# Fabrication and electrical transport characterization of high quality underdoped YBa$_2$Cu$_3$O$_{7-\delta}$ nanowires

**Eric Andersson[1], Riccardo Arpaia[1,2], Edoardo Trabaldo[1], Thilo Bauch[1] and Floriana Lombardi[1]**

[1] Quantum Device Physics Laboratory, Department of Microtechnology and Nanoscience, Chalmers University of Technology, SE-41296, Göteborg, Sweden
[2] Dipartimento di Fisica, Politecnico di Milano, I-20133 Milano, Italy

E-mail: floriana.lombardi@chalmers.se

**Abstract**

We present the fabrication and electrical transport characterization of underdoped YBa$_2$Cu$_3$O$_{7-\delta}$ nanowires. The nanowires have been realized without any protective capping layer and they show transport properties similar to those of the parent thin film, demonstrating that they have not been damaged by the nanopatterning. The current-voltage characteristics of the underdoped nanowires show large hysteretic voltage switching at the critical current, in contrast to the flux-flow like characteristics of optimally doped nanostructures, indicating the formation of a self-stabilizing hot spot. These results open up new possibilities of using the underdoped nanowires for single photon detection and for exploring the underdoped side of the YBa$_2$Cu$_3$O$_{7-\delta}$ phase diagram at the nanoscale.

Keywords: high temperature superconductors, underdoped nanostructures, superconducting detectors

## 1. Introduction

Investigating superconducting YBa$_2$Cu$_3$O$_{7-\delta}$ (YBCO) nanowires at different levels of oxygen doping can give access to a variety of new experiments to elucidate the still unknown mechanism for high critical temperature superconductors (HTS). It is, by now, well established that the complex phase diagram for HTS is dominated by various local electronic orders which are intertwined with superconductivity [1] and that become stronger in the underdoped region of the phase diagram. All recent activities on HTS nanowires are focused on optimal doping [2-7]. The realization of underdoped nanowires, with dimensions comparable to the relevant length scales of the local orders, would help clarifying how these orders are related to the high critical temperature phenomenon.

At the same time, underdoped HTS nanowires can be important for the development of quantum limited sensors where high values of resistivity could increase the sensitivity of the devices. It is well established that the resistivity of HTS thin films increases when going to the underdoped side of the phase diagram [8]. In our previous works we have demonstrated that an increased resistivity helps to improve the flux noise properties of nanowire based superconducting quantum interference devices (nanoSQUIDs) [9-11]. High resistivity is also important for superconducting nanowire based single photon detectors (SNSPDs) where it greatly enhances the sensitivity [12].

However, YBCO has some drawbacks when one is dealing with the fabrication of nanostructures. Firstly, the material is chemically unstable compared to low critical temperature superconductors (LTS), which makes it difficult to fabricate nanostructures without severe degradation [13-15]. Secondly, the reduction of the hole-doping makes underdoped nanowires much less stable against oxygen out-diffusion which further complicates the fabrication process. Currently, there are only few reports on the realization of underdoped HTS nanostructures, and the results are not very encouraging. The transport properties are severely degraded in comparison to the pristine thin films [16,17].

In this paper, we present the fabrication and electrical transport characterization of 50 nm thick underdoped YBCO nanowires without any protective capping. We demonstrate that the nanowires show properties similar to those of the pristine underdoped thin film, implying that we do not damage the material during the fabrication process. We also discuss the main physical requirements for nanowires to be employed as SNSPDs. We demonstrate that our underdoped nanowires show large hysteretic voltage switching, indicating self-sustained hot spot formation despite the comparably large thickness. These results make the nanowires promising candidates for future application as SNSPDs.

## 2. Nanowire based single-photon detectors: main requirements

In the original proposal for a SNSPD, by Golt'sman et al. [18], the absorption of a photon with energy $\hbar\omega \gg 2\Delta$ (where $\omega$ and $\Delta$ are the frequency of the impinging photon and the gap of the superconductor respectively) creates, through electron-electron and electron–phonon interactions, a local non-equilibrium perturbation with a large number of excited hot electrons. This corresponds to an increased local temperature in the nanowire, above the superconducting transition. The time for this initial thermalization phase $\tau_{th}$ must be extremely fast, in general much faster than the diffusion time $\tau_D$ of the electrons, to result in the formation of a resistive normal region hot spot which spreads through the width of the nanowire [19]. If this condition is not realized, and the thermalization time $\tau_{th}$ of the electrons is larger than their diffusion time $\tau_{D,w} \cong w^2/4D$ (where $w$ is the nanowire width and $D$ the thermal diffusion constant) across the nanowire strip, the energy of the photon will be smeared over a large area, leading to a weak influence on the superconducting properties which complicates the photon detection [19]. Thus the formation of the hot spot strongly depends on the electronic properties of the superconducting material and the dimensions of the nanowire. When searching for new materials that can be used for SNSPDs one should therefore focus on systems with low thermal conductivity, once other conditions are satisfied, e.g. a small value of the coherence length [19].

At present, the materials that best satsify these requirements, and are commonly used for SNSPDs, are granular LTS such as NbN, NbTiN and WSi [12]. An alternative route would be to use HTS [20,21], in particular YBCO nanowires [2,22-26], which could help realizing detectors that operate at higher temperatures. It is, however, challanging to use YBCO for SNSPDs. Apart from the chemical instability of the material during the fabrication process, the main issue comes from the high thermal conductivity of optimally doped YBCO ($\kappa \approx 15$ Wm$^{-1}$K$^{-1}$ in the normal state [26]) which would prevent the stabilization of a hot spot.

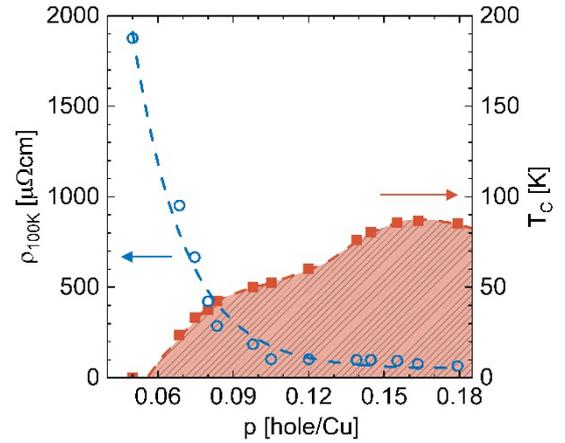

*Figure 1: Resistivity at T=100 K (blue open circles) and critical temperature $T_C$ (red filled squares) versus hole-doping level for t=30 nm YBCO thin films. The blue dashed line is an exponential fit to the resistivity data. The data is adapted from [8].*

Because of the high thermal conductivity, observation of highly hysteretic current-voltage characteristics (IVCs) in optimally doped YBCO nanowires (indicative of the formation of a self-stabilizing hot spot [28]), is hard to achieve. A possible route towards YBCO based SNSPDs is therefore to reduce the thermal conductivity by increasing the resistivity of the samples. According to the Wiedemann-Franz law, which has recently been demonstrated to hold for cuprate superconductors [29], the thermal conductivity $\kappa = LT/\rho$ (where $L = \pi^2 k_B^2/3\,e^2$ is the Lorenz number) can be reduced by increasing the resistivity $\rho$. This approach is utilized in LTS where $\rho$ is increased by introducing granularity in the thin films [30], bringing the system close to the metal insulator transition. In YBCO, a different strategy has to be adapted since grain boundary formation significantly reduces the critical current density $J_C$, of the films [31].

It has been shown, however, that hysteretic IVCs can be achieved in YBCO by maximizing the Joule heating, i.e. maximizing the critical current $I_C$ and the normal state resistance $R_N$, since Joule heating in the normal state helps to stabilize the hot spot [32]. Hysteretic IVCs have been observed in high quality (high critical current density $J_C$) YBCO nanowires that are only a few unit cells thick (high $R_N$) [5,26]. However, because of the small thickness, these nanowires have problems with stability over time and it is very challenging to fabricate long, meandering structures that are required for SNSPDs. An alternative approach to increase $\rho$ in YBCO is to reduce the hole-doping level $p$, since the resistivity increases exponentially with decreasing doping level, see figure 1. By using underdoped YBCO it might be possible to achieve hot spot formation in thicker and larger nanowires that are more stable compared to the ultrathin optimally doped ones.

## 3. Fabrication of underdoped nanowires

50 nm thick YBCO thin films were deposited by pulsed laser deposition on MgO (110) substrates at 760°C, 0.8 mbar oxygen pressure and with a laser fluency of 2 J/cm$^2$. To achieve the desired level of oxygen doping, the films were slowly cooled down in a low steady flow of oxygen resulting in a chamber pressure $P \sim 10^{-4} - 10^{-5}$ mbar, see [8] for a detailed description of the deposition process.

For the nanopatterning we use an improved process, previously developed for slightly overdoped YBCO nanowires, where the YBCO is etched with gentle Ar$^+$-ion milling through an electron-beam lithography defined amorphous carbon mask. The schematic flow of the process is shown in figure 2(a) (see [33-35] for a more detailed description). The patterned nanowires are 450 nm long and 100-600 nm wide.

To reduce the overheating and to preserve the properties of the nanowires during the Ar$^+$ etching, the sample stage was cooled with liquid nitrogen, instead of standard water cooling. Resistivity measurements of the nanowires show that with water cooling, the 200 nm wide nanowires are not superconducting while with liquid nitrogen cooling, they show a sharp transition around 75 K, see figure 2(c). Another important difference in the fabrication process, compared with our earlier works, is that the removal of the amorphous carbon mask, left on the YBCO after the Ar$^+$-ion milling, has been done with an extremely gentle reactive ion etching process. A very low power (15 W) microwave oxygen plasma is used to remove the carbon mask. For higher powers the superconducting transition of the nanowires is broadened indicating damage to the material.

## 4. Electrical transport characterization

In figure 3, the resistivity vs temperature $\rho(T)$ dependence for a nanowire and a large micro strip (from here on referenced to as "film"), are presented and compared. A first indication that the nanowires are pristine after the patterning process is that the value of the resistivity at room temperature is not significantly increased compared to the film (645 μΩcm compared to 623 μΩcm (see panel (b) and (a) of figure 3). In case of induced defects or changes in the doping due to oxygen out-diffusion from the etching, the difference in $\rho$ would have been much more significant. Moreover, from the comparison of the superconducting transition of the nanowire and the film, see figure 3(c), it is evident that the superconducting transition temperature $T_C$ is unchanged and there is very little additional broadening of the transition induced by the nanowire.

Since our underdoped nanowires are "bare YBCO", i.e. without any gold capping layer, one can also study the normal state transport properties of the material. The YBCO (and HTS in general) phase diagram is very rich in the underdoped side with various local orders simultaneously at play [1]. Signatures of variation of the Fermi surface topology and transport mechanisms are also visible in the electronic transport. An interesting question is if these orders are also present in nanowires of reduced dimensions. In figure 3, the characteristic temperatures $T^*$ and $T^{**}$ are pointed out. The pseudogap temperature $T^*$ signifies the crossover from the strange metal phase ($\rho \propto T$) [36] to the pseudogap phase in HTS (connected to the breaking of the Fermi surface and the formation of Fermi arcs [37] at $T^*$) in the underdoped regime. The temperature $T^{**}$, which is lower than $T^*$, instead indicates a crossover to more Fermi-liquid like ($\rho \propto T^2$) transport behaviour. $T^*$ and $T^{**}$ does not change significantly when reducing the dimensions (see figure 3(a) and 3(b)). Hence the physical properties of the film and the nanowire are the same.

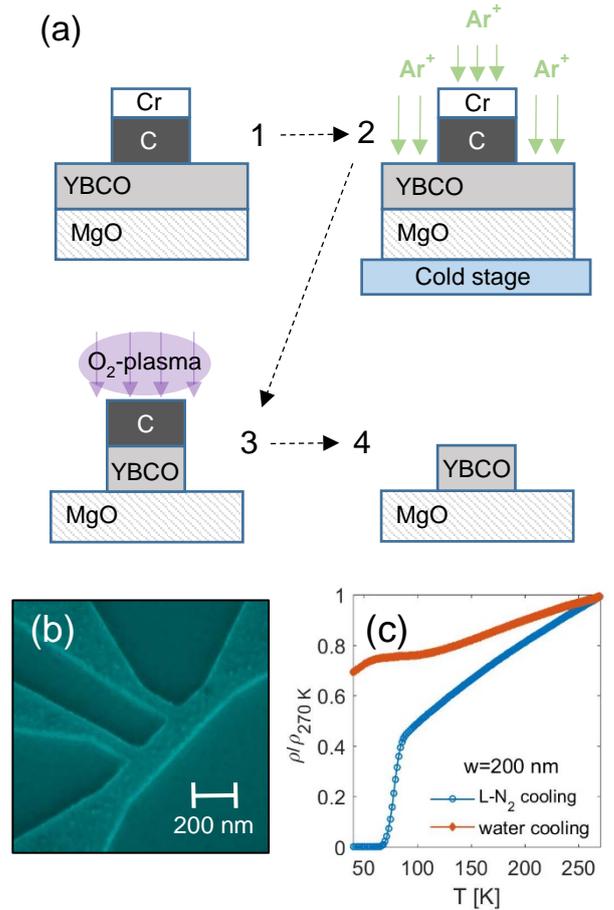

Figure 2. (a) Illustration of the nanowire fabrication process. In step 1, a hard C mask has been deposited and patterned on the YBCO thin film by $O_2$ plasma etching through an electron-beam lithography defined Cr mask. In step 2, the YBCO not covered by C and the Cr mask is removed with Ar$^+$ ion milling. In 3, the remaining C on top of the YBCO is etched with a low power $O_2$ plasma. The final result is shown in 4. (b) False-colour SEM image of a 200 nm wide nanowire with current and voltage leads. (c) Resistivity vs temperature measurement of underdoped nanowires etched with water cooling (red filled diamonds) and liquid nitrogen (L-$N_2$) cooling (blue open circles) of the cold stage during the ion milling in step 2 (panel (a)). Note that only the nanowire etched with L-$N_2$ shows a superconducting transition.

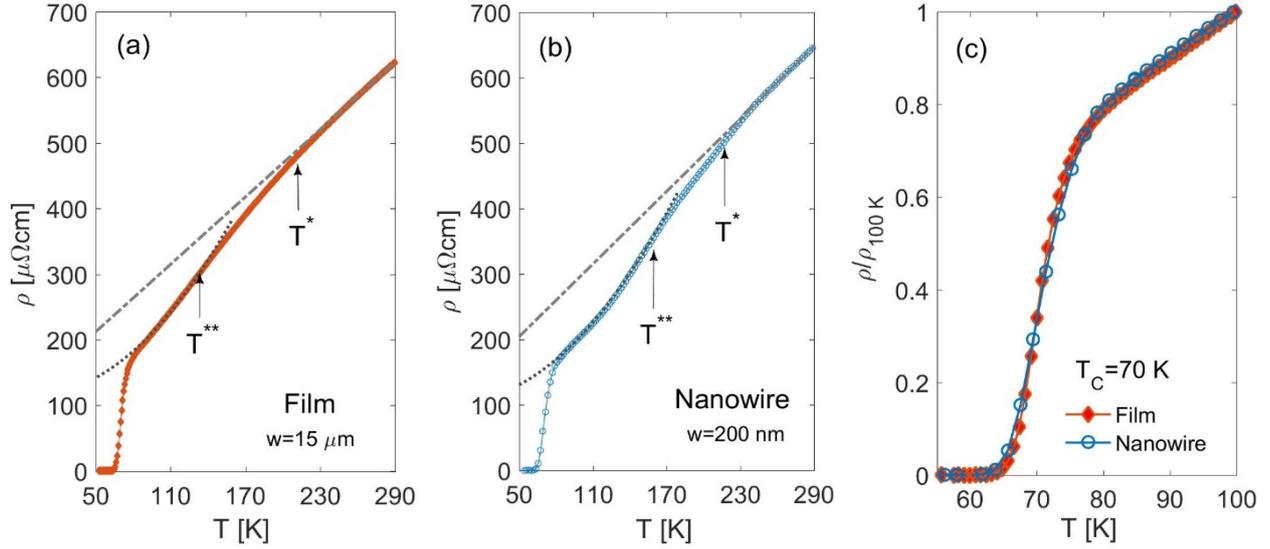

*Figure 3: Resistivity versus temperature data for (a) 15μm wide strip representing the underdoped film properties ($T_c$ =70 K), (b) a w=200 nm wide nanowire and (c) comparison of the wide strip and nanowire resistivity normalized at T=100 K. Panel (c) clearly shows that the nanopatterning process does not significantly affect the superconducting transition. The characteristic temperatures $T^*$ and $T^{**}$ are highlighted for comparison in (a) and (b). $T^*$ is the temperature where the resistivity deviates from high temperature linear behaviour and $T^{**}$ the onset of quadratic resistivity behaviour.*

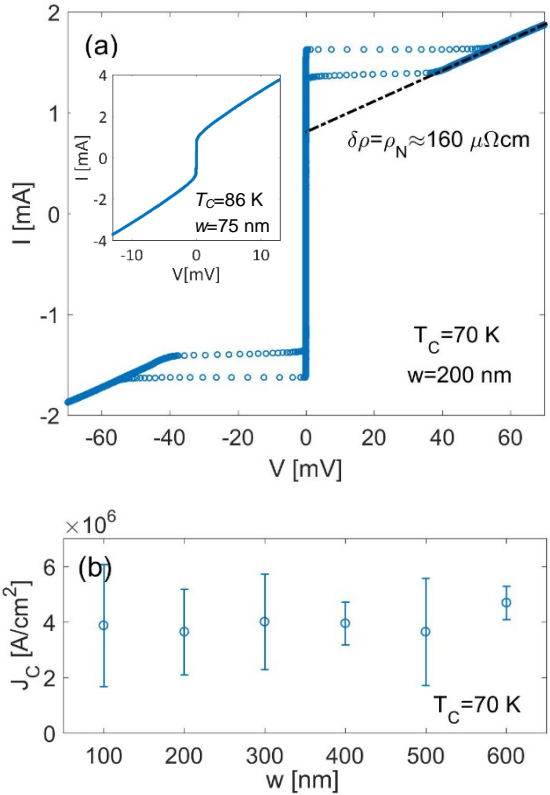

*Figure 4: (a) typical IVC of a 50 nm thick underdoped nanowire at T=4.5 K and (b) width dependence of the critical current density $J_C$. Each point represent data from four nanowires patterned on the same thin film. The inset in (a) shows a typical IVC of a slightly overdoped nanowire.*

Figure 4(a) shows the IVC for a typical underdoped nanowire ($p \approx 0.13$) measured at $T$=4.5 K. It shows large hysteretic voltage switching at the critical current $I_C$, indicative of a self-sustained hot spot [28]. For comparison, a typical IVC of a slightly overdoped YBCO nanowire is shown in the inset of figure 4(a). For overdoped nanowires, the IVC is flux-flow like, without any sign of voltage switching, in spite of the narrower width ($w$=75 nm). The difference has to be attributed to the higher resistivity in the underdoped case, leading to a reduced thermal conductivity which promotes Joule heating and the self-stabilization of the hot spot. Another fact that supports this hypothesis is that only the nanowires with the highest $J_C$ (where Joule heating is maximized) show voltage switching, while the nanowires with low $J_C$ show flux-flow behaviour. In the overdoped regime, it is still possible to observe hysteretic voltage switching if the dimensions are reduced to t=10 nm and w<100 nm [26]. Such extreme dimensions result in instability of the nanowire electrical properties over long periods of time.

The dynamical resistivity, $\delta\rho = dI/dV \times A/l$ where $A$ is the cross-sectional area of the nanowire and $l$ the length, is extracted from the slope of the IVC in the normal state (see figure 4(a)). The value $\delta\rho$ =160 μΩcm (corresponding to $R \approx$ 76 Ω) is close to the normal resistivity $\rho_N$ of the nanowire just above the superconducting transition (see figure 3(b)). This indicates that the whole nanowire volume turns resistive after the switch, proving effective spreading of the hot spot.

Since a high value of $I_C$ is important for the voltage switching to occur in the nanowires [26], we have measured the critical current density $J_C$ dependence as a function of the width (see figure 4(b)). $J_C$ is approximately constant with width down to $w$ =100 nm which enables the realization of even narrower structures in the future. The value of the critical current density $J_C \approx 4 \times 10^6$ A/cm$^2$ is only half of the best values measured in underdoped YBCO nanowires capped with Au [8], which indicates that the Au capping is indispensable for achieving the highest $J_C$.

## 5. Conclusions

We have presented the realization of high-quality underdoped YBCO nanowires. They are without a metallic protective capping and are therefore suitable for single photon detection and for studying the normal state of HTS on the nanoscale. The doping level of the nanostructures is the same as that of the parent thin film with transport properties unaffected by the fabrication process. Despite the relatively large thickness ($t$=50 nm), the IVCs show large hysteretic voltage switching that in optimally doped-overdoped YBCO nanowires can only be achieved at extreme thicknesses (<10 nm) and widths (<100 nm). This demonstrates that for underdoped nanowires the increase in resistivity relates to a reduction of the thermal conductivity in agreement with the Wiedemann-Franz law. The large hysteretic voltage switching makes underdoped nanowires attractive for application as SNSPDs. There are, however, some minor problems to address. The normal state resistance of the nanowires after the switch is not very large ($R \approx 76\Omega$) which would limit the sensitivity in a SNSPD setup. To increase the resistance, one possibility is to reduce the thickness to values which are more common for SNSPDs ($t \approx$ 5-10 nm) and/or increase the nanowire length. This, however, represents a challenge since thinner and longer wires would be more unstable over time and suffer from a significant reduction of $J_C$ [24]. Another way of increasing the resistivity is to reduce the doping even further. Here we have presented devices with $T_C$ around 70 K. In principle there should be no major issues in making nanowires with $T_C$ in the range of 50 K with properties comparable to the ones presented in this paper.

**Acknowledgements**

This work was supported in part by the Knut and Alice Wallenberg Foundation (KAW) and in part by the Swedish Research Council (VR). R.A is supported by VR under the project 2017-00382.